\documentclass{PoS}
\usepackage{graphicx}

\newcommand{\be}{\begin{equation}}
\newcommand{\ee}{\end{equation}}
\newcommand{\bea}{\begin{eqnarray}}
\newcommand{\eea}{\end{eqnarray}}
\newcommand{\pslash}{\not\!p}

\title{Strong Coupling QCD in the Continuum}
\ShortTitle{Strong Coupling QCD}

\author{\speaker{M.~R. Pennington}\\{Institute for Particle Physics Phenomenology, Physics Department,\\ Durham University, 
        Durham DH1 3LE, U.K.}\\
E-mail: \email{m.r.pennington@durham.ac.uk}
         }

\abstract
{A short overview of studies of strong coupling problems in hadron physics is presented.
The Schwinger-Dyson/Bethe-Salpeter approach in the continuum allows the perturbative regime to be connected to the world of strong coupling that controls  confinement and chiral dynamics that in turn determines the properties of light hadrons with just one scale, {\it viz.} $\Lambda_{QCD}$, and with the addition of current masses the properties of heavy flavours too.}

\FullConference{International Workshop on Effective Field Theories: from the pion to the upsilon \\
		February 2-6 2009\\
		Valencia, Spain}

\begin{document}

\section{QCD in the continuum}
QCD is a remarkable theory. At short distances, thanks to asymptotic freedom,
quarks move as though they are free. There probed in hard scattering processes we can use perturbation theory, make predictions and find that these agree with experiment. However, at longer distances, distance scales of the size of a hadron, the interaction becomes strong. This is the region that controls confinement, the dynamical generation of mass and the binding of quarks to make hadrons~\cite{swimming,craig1}.

To illustrate the methods and how they work, let us consider the problem of mass generation~\cite{miransky,swimming}. We know that at short distances, the {\it up} and {\it down} quarks not only behave as though they barely interact, but are almost massless too. What we have learnt is that over longer distances strong dynamics creates correlations in the vacuum that generates a mass of 300 MeV, dressing the quarks to become ``constituent'' quarks. We, of course, know that such dynamical generation of mass must be a strong physics phenomenon, because in perturbation theory if a bare mass is zero, then the mass is zero to all orders in the coupling. Moreover, since zero mass particles will not fit on a finite size lattice, we know we have to study how masses are generated in the continuum.
To understand when this can happen, we consider the Schwinger-Dyson Equation (SDE) for the fermion propagator in a gauge theory.

 The SDEs are the field equations of the theory, Fig.~1. 
The fermion propagator, $S_F(p)$, in momentum space depends on two functions, its wavefunction renormalization $F(p)$ and its mass function, $M(p)$:  
\be
S_F(p)\;=\;\frac{F(p)}{\pslash\,-\,M(p)}\quad .
\ee
Solving the SDE for the fermion, Fig.~1, requires knowledge of the boson propagator
and the full fermion-boson interaction. In QED, where the equations are simpler to draw as in Fig.~1, the photon propagator depends on the very same full fermion propagator and fermion-boson vertex. So one can imagine solving the coupled equations for the two propagators (or 2-point functions), provided we know the 3-point vertex. But this in turn satisfies an SDE, again shown in Fig.~1, that relates the 3-point function to the 2, 3 and 4-point functions, and the 4-point function is related to the 2,~3,~4 and 5-point functions, and so on {\it ad infinitum}. Consequently, we have an infinite tower of coupled equations that we cannot solve
without some truncation. The best known approximation is to expand each Green's function in powers of the coupling. This gives us perturbation theory, which  satisfies 
gauge invariance and multiplicative renormalizability at each order of truncation. While both the fermion and boson equations each involve two functions (as in Eq.~1.1), the 3-point equation contains 12 equations, and the tower becomes increasingly complex. 

To see how to proceed, let us start by butchering the fermion equation. We can cut this off from the ``tower'' by treating the photon propagator and the vertex as bare, leaving just the fermion propagator as dressed.
We can then perform the angular integrals and project out the functions
$F$ and $M$ to obtain~\cite{swimming} in any covariant gauge $\xi$:
\begin{figure}[t]
\begin{center}
\includegraphics[width=0.75\textwidth]{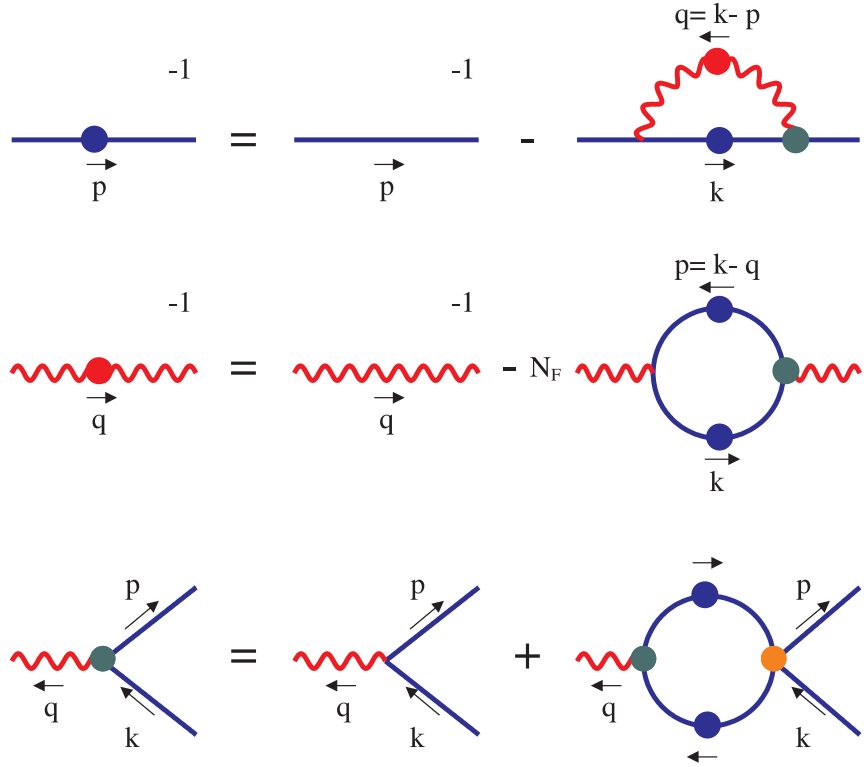}
\caption{Schwinger-Dyson equations in QED for the fermion and boson propagators and their interaction. The solid dots denote full Green's functions.}
\end{center}
\end{figure}
\bea
\frac{M(p)}{F(p)}&=&m_0\;+\;\frac{\alpha_0}{4\pi}\, (3+\xi)\,
\left[ \int_0^{p^2}\,dk^2\;\frac{k^2}{p^2}\;\frac{F(k)\,M(k)}{k^2\,+\,M(k)^2}\,+\,\int_{p^2}^{\kappa^2}\,dk^2\;\frac{F(k)\,M(k)}{k^2\,+\,M(k)^2}\right]\quad ,\\
\frac{1}{F(p)}&=&1\;+\;\frac{\alpha_0\,\xi}{4\pi}\, \,
\left[ \int_0^{p^2}\,dk^2\;\frac{k^4}{p^4}\;\frac{F(k)}{k^2\,+\,M(k)^2}\,+\,\int_{p^2}^{\kappa^2}\,dk^2\;\frac{F(k)}{k^2\,+\,M(k)^2}\right]\quad.
\eea
Of course, this is not a calculation of a realistic situation. The photon being bare, the coupling, $\alpha_0$, does not run and there is no intrinsic scale. Consequently, in Eqs.~(1.2, 1.3) we have introduced a cutoff $\kappa$ on the integrals.
 This sets the scale prior to renomalization.
Looking at these equations we see that, if we work in the Landau gauge with $\xi=0$, Eq.~(1.3) simplifies further with $F =1$. This leaves the mass equation, Eq.~(1.2), which always has the solution $M= 0$ if the bare mass $m_0=0$. But when can a non-zero mass function be generated? If the coupling is big enough, a mass function that looks qualitatively similar to those in QCD, shown in Fig.~2, can indeed be found: non-zero in the infrared  and then dying away at larger momenta. Such solutions only happen if the fixed coupling $\alpha_0 \ge \pi/3$. Thus, if the interaction is strong enough, a  mass can be generated. However, if we now solve Eqs.~(1.2, 1.3) in other covariant gauges, we see the same qualitative behaviour, but this happens at different values of the coupling. This tells us our approximation, in particular the use of a bare interaction, does not respect gauge invariance. Of course, the Ward-Green-Takahashi identity (WGTI)~\cite{wgt}, which relates a projection of the fully dressed 3-point function to the inverse of the fermion propagator, is a key consequence of gauge invariance that the bare vertex does not satisfy, except in massless, quenched QED. What we have learnt from two decades of study~\cite{swimming} is that imposing the WGTI together with the requirement of multiplicative renormalizability pulls through from the tower of SDEs just what is required to obtain physically meaningful results.

 This is the technology that we take to the study of QCD, together with the expectation that dynamical mass generation will occur when the coupling is strong.

\section{First studies in strong coupling QCD}
  
The field equations of QCD are of course more complicated than QED, involving not only the self-interaction of gluons, but in covariant gauges also ghosts. An important Slavnov-Taylor identity~\cite{STI} is that relating the 3-gluon interaction to the inverse propagator of the gluon. In axial gauges this is particularly simple, and was the starting point for the analysis of Baker, Ball and Zachariasen (BBZ)~\cite{bbz}. Though in axial gauges there are no ghosts, the gluon propagator depends on two independent functions. BBZ made an approximation that only one of these is relevant, but this was shown by West~\cite{west} to be inconsistent in the infrared region. Consequently, attention turned to covariant gauges, in particular the Landau gauge, where
the gluon propagator depends on just one function $\Delta(q^2)$. The corresponding Slavnov-Taylor identity for the dressed 3-gluon interaction now involves ghost functions.
The first studies 25-30 years ago set these to one, and assumed that the role of ghosts was merely to ensure the physical gluons were transverse, but otherwise their effect was negligible/neglectable.
Only gluon dressing was deemed required for confinement dynamics. These studies started independently with Pagels~\cite{pagels}, Mandelstam~\cite{mandelstam} and Bar-Gadda~\cite{bargadda}, and indicated that the gluon propagator became singular in the infrared.
\begin{figure}[b]
\begin{center}
\includegraphics[width=0.65\textwidth]{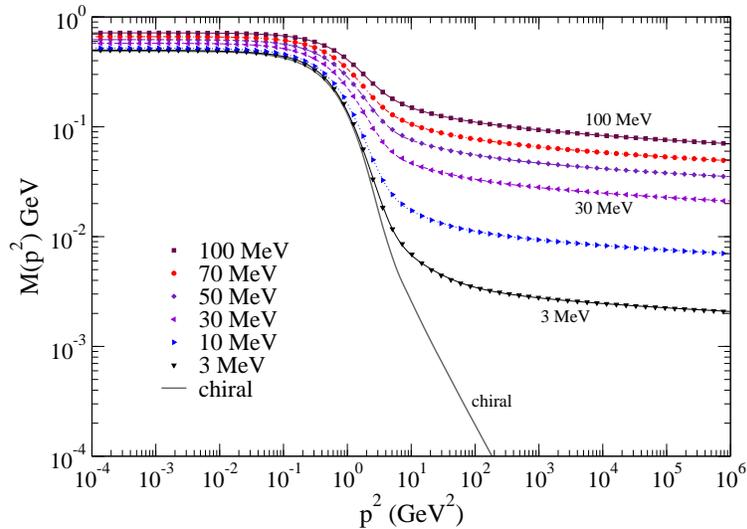}
\caption{Euclidean mass functions labelled by their current masses specified at $\mu=19$ GeV covering the range appropriate for the 3 lightest flavours of quark and the chiral limit. 
These results
 from Ref.~\cite{williams} are essentially the same as those obtained earlier by Maris and Roberts~\cite{MR} covering a bigger range of masses required for 5 flavours, relevant to the discussion in Sect.~3.}
\end{center}
\vspace{-3mm}
\end{figure}
The first major numerical study was performed by Nick Brown and myself~\cite{brown} more than twenty years ago, which showed how the gluon enhancement was correlated with
$\Lambda_{QCD}$, and that the addition of generations of massless quarks softened this enhancement. The strong coupling was enhanced in the infrared, but merged with standard perturbation theory above 
a few GeV. Such behaviour generates dynamical masses for the {\it up} and {\it down} quarks, as shown by Maris and Roberts~\cite{MR}. In Fig.~2 is plotted the behaviour of the quark mass function for a range of current masses~\cite{williams} from 3 to 100 MeV. In each case the \lq\lq current'' quarks are dressed by roughly 300 MeV of mass at infrared momenta to emerge as \lq\lq constituent'' quarks, the scale being set by $\Lambda_{QCD}$. Having these solutions in the continuum, we can take the current mass to zero
and so determine the chiral limit. Then  matching the behaviour in Fig.~2 to the Operator Product Expansion at large momenta, we learn this corresponds to a non-zero chiral condensate with $\langle\, {\overline q} q\,\rangle_0\;=\;- (240\, {\rm MeV})^3$~\cite{MR}, in agreement with experiments on low energy $\pi\pi$ scattering~\cite{gasser,na48}.

For later, we note that the results for the quark mass functions $M(p)$ of Fig.~2 and their extension to GeV current masses~\cite{MR}, can be parametrised by
\be
M(p) \simeq M_0\,+\,c\, \frac{\Lambda_{QCD}^3}{p^2+\Lambda_{QCD}^2}\quad ,
\ee
if we simply ignore the anomalous dimensions. 
$M_0$ is the current mass,  $c$ a dimensionless constant fixed from Fig.~2, and
where $\Lambda_{QCD}$ sets the scale for the infrared dressing and the ${\overline q}q$ condensate.

How is an infrared enhanced gluon related to confinement? The interaction between quarks and antiquarks is given by an infinite set of gluon exchange graphs. However, if we consider the infinite quark mass limit, then all those with internal quark propagators are suppressed and the quark-gluon vertex becomes bare. This is the basis of ``Heavy Quark Effective Theory''. Then the interquark potential is generated by one gluon exchange, and the static potential $V(r)$ in position space is just the Fourier transform of the time-time component of the dressed gluon propagator, $\Delta(q)$. This generates a wholly vector potential and so must be modified for physical, finite mass quark effects. For heavy quarks, it is just a matter of dimensional analysis to see that the behaviours
\be
V(r)\;\sim\;r^a \qquad {\rm and} \qquad \Delta(q)\;\sim\; q^{-3-a}
\ee
are correlated when $q\, r \sim 1$. This naturally accords with expectation
at large momenta when the gluon is essentially free, behaving like $1/q^2$, {\it ~i.e.} $a=-1$. This corresponds to a Coulomb-like potential at short distances. If at larger distances the potential is to grow linearly with $r$, then this static approximation requires $\Delta(q) \sim 1/q^4$ for $q \ll 1$. Neglecting ghosts this is just what happens.

\section{T\"ubingen/Graz/Darmstadt studies of QCD}

\begin{figure}[t]
\begin{center}
\includegraphics[width=0.96\textwidth]{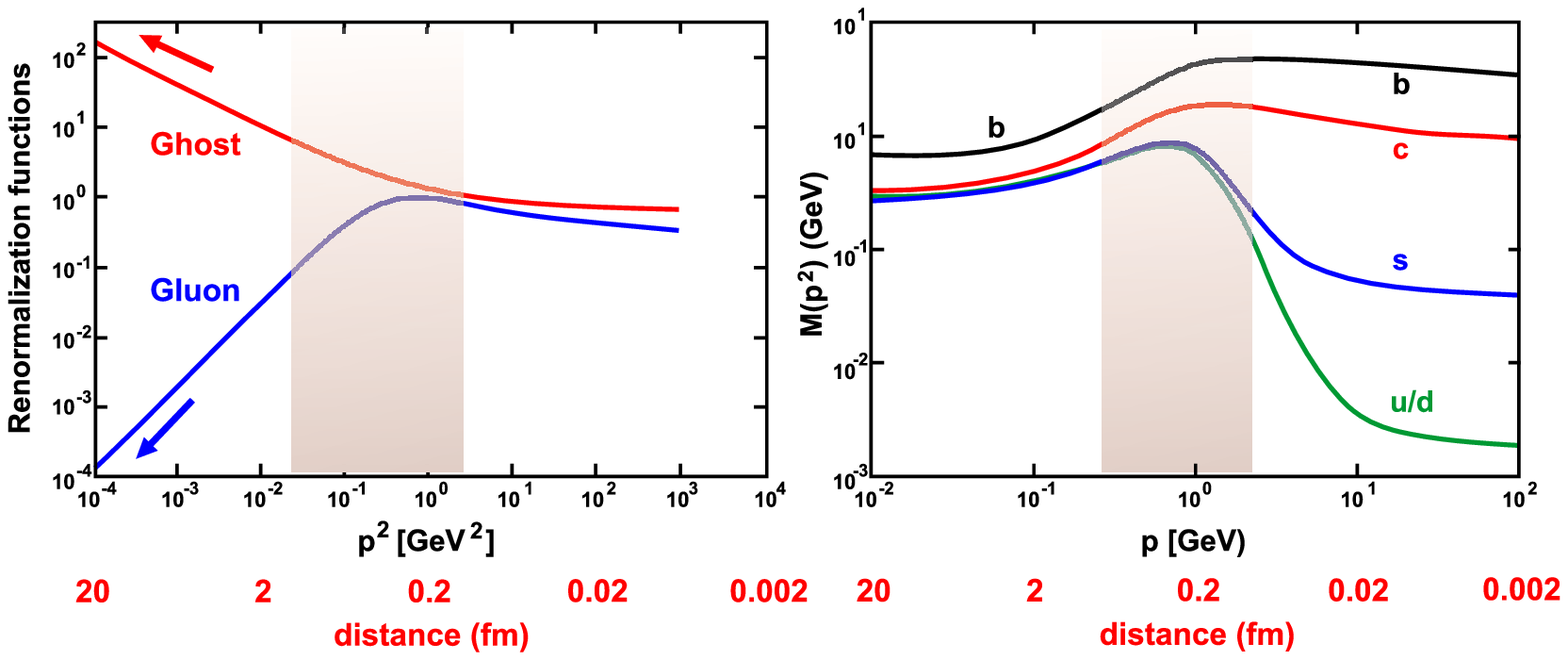}
\caption{Log-log plots of (a) the gluon and ghost renormalization functions, and (b) the quark mass functions from the studies in Refs.~\cite{fischer,llanes-estrada}.}
\end{center}
\end{figure}
The next group to pursue intensive studies started in T\"ubingen, but is now dispersed all over the globe, with von Smekal, Alkofer, Fischer and collaborators~\cite{lorenz,fischer}. They showed that ghosts {\underline {do}} play an important role. It is ghosts that become enhanced in the infrared, while the gluon propagator with its physical transverse polarisations is suppressed.  In this \lq\lq scaling'' solution, the behaviour of the ghost and gluon are correlated, Fig.~3, so
that the effective coupling becomes enhanced with a finite value in the infrared. A major effort has been undertaken to check the consistency of the tower of SDEs in the deep infrared, {\it e.g.}~\cite{schwenzer}. While mathematically well defined these studies probe the behaviour of the Green's functions when the hadron universe is more than 10 fermis large, Fig.~3. However, the physics of hadrons only depends on QCD dynamics from 0.1 to at most 2 fermis. There the enhancement of the coupling is expected to lead to dynamical chiral symmetry breaking, as in the studies where ghosts were neglected. This has been shown explicitly in Ref.~\cite{llanes-estrada}. Key to these studies is the non-trivial structure of the quark-gluon interaction with a crucial scalar component generated by the contribution of the ghost loop in Fig.~4. In the limit in which the two fermion momenta, $p, p'$, are small, one would expect the contribution of such a graph to depend on the quark mass as sketched in Eq.~(3.1):
\begin{figure}[h]
\begin{center}
\includegraphics[width=1.0\textwidth]{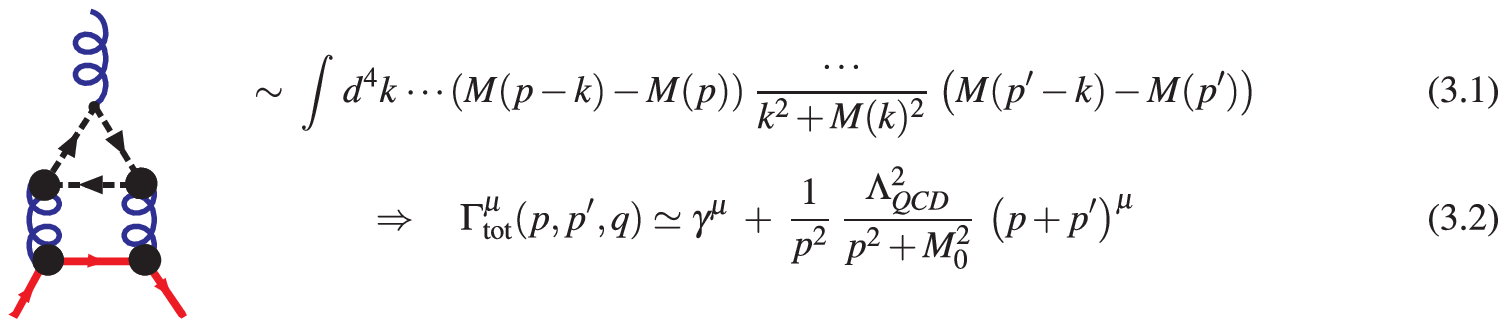}
\caption{The ghost triangle claimed in Ref.~\cite{llanes-estrada} to enhance the quark-gluon vertex even in the heavy quark limit.}
\end{center}
\vspace{-7mm}
\end{figure}

\noindent
If the quark mass function behaves for heavy quarks as illustrated in Fig.~2 and given by Eq.~(2.1), then the corrections, Fig.~4, to the bare vertex would be like those  in Eq.~(3.2), and so be suppressed in the infinite quark mass limit. However, Alkofer {\it et al.}~\cite{llanes-estrada} find their quark mass functions have a rather different behaviour --- as shown on the right in Fig.~3.  While the quarks all gain 300 MeV of mass above their {\it current} value in the phenomenologically important region of momenta of 0.1 to 1 GeV, the heavy quark masses decrease in the deeper infrared. This behaviour circumvents the underpinning assumptions of the heavy quark limit discussed above. Remarkably, Alkofer {\it et al.}~\cite{llanes-estrada} then find that a potential growing linearly with distance emerges.
I personally find the low momentum behaviour of the ``heavy'' quark mass function of Fig.~3 rather bizarre, and demands further investigation. However, we do see that in the region probed by hadron physics that this modelling is just as likely as that with the enhanced gluon. This will be confirmed by calculation of hadron masses.



As yet another alternative, Mandula and Ogilvie \cite{MO} long ago proposed a solution for the gluon with an effective mass, and this has been more recently studied by Aguilar {\it et al.}~\cite{aguilar}. Such behaviour appears to be correlated with a finite infrared enhancement of the ghost, as found by Boucaud {\it et al.}~\cite{paris} in particular. Interestingly, this type of solution appears to be in qualitative agreement with recent lattice results, for instance those of Ref.~\cite{cucchieri}. However, these SDE solutions have been dismissed by the proponents of the \lq\lq scaling'' solution~\cite{lorenz}-\cite{llanes-estrada} in Ref.~\cite{pawlowski}.

An aim of this whole approach is to build hadrons and to be able to do this for physical mass quarks~\cite{tandy,craig1}. Something lattice computations are just being able to study. Knowing the quark and gluon propagators and their interactions allows us to solve the Bound State (or Bethe-Salpeter) equations (BSE). We start with ${\overline q}q$
systems. The key dynamics is provided by the quark scattering kernel and most studies use the rainbow ladder approximation~\cite{tandy}. For states with pseudoscalar quantum numbers satisfying the axial Ward identity is crucial~\cite{maris,craig2}. This ensures that in the limit of zero mass quarks, the pion is massless with chiral symmetry dynamically broken.  Moreover, pion-pion interactions have much of the essence of Chiral Perturbation Theory: not all, as the kernels do not yet include a complete pseudoscalar nonet, but this approach is getting there. The masses of scalars are found to be  sensitive to the details of the scattering kernel. Consequently, the shape of scalar-pseudoscalar potential cannot yet be mapped out with any confidence.
However, vector bound states are more robust. The $\rho$ and $\pi$ masses can be evaluated for a whole range of quark masses with interactions modelled to reproduce the enhanced gluon scenario or the suppressed gluon one~\cite{tandy,benhaddou,graz}, and both  match the lattice results of CP-PACS~\cite{pacs}. This highlights how the SDE/BSE approach can continue lattice results with their unphysical current quark masses to those of the real world rather well, and that the SDE/BSE system encodes the
key strong coupling dynamics that extends QCD perturbation theory into the confinement  regime --- and that this does not depend critically on knowledge of the deep infrared.

One can of course use this same approach to study other dynamical quantities. Electromagnetic form-factors are naturally first~\cite{craig3}, where a simple impulse approximation may be justifiable. The spacelike pion form-factor is the benchmark for the
extensions of perturbation theory pioneered by Brodsky and Lepage~\cite{brodsky}. Studies in the SDE/BSE approach reproduce experiment and show the perturbative limit is not reached before $Q^2 \simeq 50$ GeV$^2$. More recently attention has turned to the properties of baryons~\cite{craig3}, an even more challenging problem requiring a full Fadeev treatment. However, we can proceed more simply, and perhaps more profitably, by noting that having forms for the quark and gluon Green's functions allows us to study not just the BSE for quarks and antiquarks, but also for diquarks.       
Baryons are then built of diquark-quark systems, and measuring their electromagnetic transition form-factors with precision at JLab will probe the strong coupling regime and the predictions of the SDE/BSE treatment of QCD~\cite{jlab}.
Such an interplay between theory and experiment promises instructive insight into how confinement in QCD really works.

\vspace{3mm}
The author acknowledges partial support of the EU-RTN Programme, Contract No. MRTN--CT-2006-035482, \lq\lq Flavianet'' for this work.

\vspace{0.5mm}


\end{document}